\def\Granadainst{Instituto de F\'\i sica Te\'orica y Computacional Carlos I,
Facultad de Ciencias, Universidad de Granada, Campus de Fuentenueva,
Granada 18002, Spain}

\def\IAA{Instituto de Astrof\'{\i}sica de Andaluc\'{\i}a, Apartado Postal 3004,
18080 Granada, Spain}

\def\Murcia{Departamento de Matem\'atica Aplicada, Facultad de Inform\'atica, Campus
de Espinardo, 30100 Murcia, Spain}

\def\nn{\nonumber}
\def\ni{\noindent}

\def\be{\begin{equation}}
\def\ee{\end{equation}}
\def\bea{\begin{eqnarray}}
\def\eea{\end{eqnarray}}
\def\ba{\begin{array}}
\def\ea{\end{array}}

\def\z{\zeta}

\def\Gt{$\widetilde{G}\,\,$}
\def\Gtm{\widetilde{G}\,}

\def\calG{ {\cal G} }
\def\calH{ {\cal H} }
\def\calP{ {\cal P} }
\def\Rad{ {\rm Rad} }

\def\z{\zeta}
\def\medio{\frac{1}{2}}

\def\ThetaPC{\Theta_{PC}}

\def\ni{\noindent}

\def\lo{\lambda^0}
\def\vlo{\vec{\lambda}^0}

\def\Real{{\mathbb R}}

\newcommand{\parcial}[1]{ \frac{\partial}{\partial #1} }

\newcommand{\derivparcial}[2]{ \frac{\partial #1}{\partial #2} }

\newcommand{\XL}[1]{ {\tilde{X}}^{L}_{#1} }

\newcommand{\XR}[1]{ {\tilde{X}}^{R}_{#1} }

\documentclass[12pt]{article}
\usepackage{amsfonts,amssymb}

\begin{document}



\begin{center}
{\large {\bf INVARIANT MEASURES ON POLARIZED SUBMANIFOLDS IN GROUP
QUANTIZATION}}
\end{center}

\bigskip
\bigskip

\centerline{J. Guerrero$^{1,2,3}$ and V. Aldaya$^{2,3}$}

\bigskip
\centerline{November 20, 1999}
\bigskip

\footnotetext[1]{\Murcia}
\footnotetext[2]{\IAA}
\footnotetext[3]{\Granadainst}

\bigskip

\begin{center}
{\bf Abstract}
\end{center}

\small

\begin{list}{}{\setlength{\leftmargin}{3pc}\setlength{\rightmargin}{3pc}}
\item We provide an explicit construction of quasi-invariant measures on
polarized coadjoint orbits of a Lie group $G$.
The use of specific (trivial) central
extensions of $G$ by the multiplicative group ${\ R}^+$ allows us to restore
the strict invariance of the measures and, accordingly, the unitarity of
the quantization of coadjoint orbits. As an example, the representations
of $SL(2,\Real)$ are recovered.
\end{list}

\normalsize

\vskip 0.25cm





\section{Introduction}

The aim of this paper is to proceed a bit further in search of a
unified algorithm for achieving unitary and irreducible representations
(unirreps for short) of Lie groups in the context of quantization. Our
starting point here is a rather developed Group Approach to Quantization (GAQ)
(see \cite{JMP23,Comm1,frachall} and references there in), which generalizes
and improves Geometric Quantization (GQ) and/or the Coadjoint-Orbit Method
(COM) \cite{Kirillov,Kostant} in many respects, and particularly in the
treatment of the non-K\"ahler orbits of the Virasoro group \cite{virazorro},
denominated ``non-quantizable orbits'' in Ref. \cite{Witten}.

GAQ inherited, however, the technical problem of finding an appropriate and
natural integration measure on the polarized submanifold of the original
symplectic coadjoint orbits (or classical phase space). In fact, even though
a symplectic manifold $(M^{2n},\;\omega)$ is canonicaly endowed with a volume,
that is, $\omega^n$, a maximally isotropic submanifold associated with a
Polarization (half a symplectic manifold, so to speak) does not necessarily
possess a canonical measure invariant under the action of the group generators
(or quantum operators, in physical language).

Nevertheless, the virtue of GAQ  working directly on a group manifold,
rather than on a coadjoint orbit, taking advantage of the tools available on 
any Lie group (left- and right-invariant vector fields, Haar measure, etc.) 
brings out again the solution to the present problem of finding invariant 
measures. The precise technique of pseudo-extensions employed here was 
introduced in \cite{pseudo} on an equal footing with non-trivial central 
extensions, and was further elaborated in \cite{Marmo}, emphasizing its 
relation with COM. Now the main
trick consists in considering pseudo-extensions by the multiplicative real line
$\Real^+$ along with (pseudo)-extensions by $U(1)$. Central (even trivial)
extensions by $\Real^+$ can modify well the common factor accompanying the wave
functions (the weight) with an extra non-unimodular real function, thus
providing half of the correction needed to make a quasi-invariant measure 
strictly invariant. The resulting construction shed new light on the
cryptic language of ``half-forms'' \cite{Woodhouse}, which came to faint the
beauty of the original scheme of GQ.

This paper is organized as follows. In Sec. \ref{pseudoextensiones} we
provide a general background on pseudo-extensions and the explicit connection
with the coadjoint orbits of a general simply connected Lie group. In Sec.
\ref{quasi-invariant} the existence and uniqueness of a quasi-invariant
measure $\mu$ with Radon-Nikodym derivative $\lambda$  on a homogeneous space
is translated to the group $G$ itself, providing a constructive proof of the
existence of such a $\lambda$. Then, with the aid of this function, we find a
specific $\Real^+$ pseudo-extension of $G$ making $\mu$ strictly invariant.
The results above are applied, as an example, to the explicit construction of
the representations of $SL(2,\Real)$, including the Mock representation.

\section{Pseudo-extensions}
\label{pseudoextensiones}

 A pseudo-extension of a simply connected Lie group $G$ is a central extension
\Gt of $G$ by $U(1)$
by means of a 2-cocycle\footnote{We shall consider, following Bargmann
\cite{Bargmann}, local exponents $\xi:\ G\times G\rightarrow R$
such that $\omega=e^{i\xi}$ defines a 2-cocycle (or factor),
$w:\ G\times G \rightarrow U(1)$.} $\xi_\lambda: G\times G \rightarrow R$,
which is a coboundary and therefore defines a trivial central extension;
i.e. there
exists a function $\lambda: G \rightarrow R$, the generating function of the
coboundary, such that $\xi_\lambda(g',g)=\lambda(g'*g)-\lambda(g')-
\lambda(g)$, but with the property that the Lie derivative of $\lambda$ at
the identity is different from zero for some left-invariant vector fields.
In other words, the gradient of $\lambda$ at the identity,
$\lambda^0_i\equiv \derivparcial{\lambda(g)}{g^i}$, with respect to a basis
of local canonical coordinates $\{g^i\}$ at a neighbourhood of the identity
of $G$, is not zero.

It should be emphasized that $\vec{\lambda}^0\equiv (\lambda^0_1,\ldots,
\lambda^0_n)$ defines an element in the dual $\calG^*$ of the Lie algebra
$\calG$ of $G$. Before going further into the properties of pseudo-extensions
and their classification into equivalence classes (in the same way as true
extensions), we must introduce some definitions.

Let $\{X^L_i\}$ be a basis of left-invariant vector
fields associated with the canonical coordinates  $\{g^i\},\, i=1,\ldots,
n={\rm dim}G$ at the identity. Let $\{\theta^{L\,i}\}$ be the dual basis of 
left invariant 1-forms on $G$. They verify the relations:
\begin{eqnarray}
i_{X^L_i}\theta^{L\,j} &=& \delta^j_i \nn \\
L_{X^L_i}\theta^{L\,j} &=& C_{ik}^j \theta^{L\,k}\,, \label{thetaproperties}
\end{eqnarray}

\ni where $C_{ik}^j$ are the structure constants of the Lie algebra $\calG$
generated by $\{X^L_i\}$.

 Right-invariant vector fields $\{X^R_i\}$ can also be introduced together 
with the dual
basis of right-invariant 1-forms $\{\theta^{R(i)}\}$, satisfying properties
similar to (\ref{thetaproperties}), but changing $C_{ik}^j$ by $-C_{ik}^j$,
since right-invariant vector fields generate an algebra isomorphic to that
of left-invariant ones but with the structure constant with opposite
sign\footnote{This is due to our choice for the left and right action of
the group on functions: $R_{g'}f(g)=f(g*g')$ and $L_{g'}f(g)=f(g'*g)$
instead of $L_{g'}f(g)=f(g'{}^{-1}*g)$, as is used in other contexts.}.
Left-invariant 1-forms
have zero Lie derivative with respect to right-invariant vector fields and
vice versa, as it should be. An important formula which will be extensively
used in this paper is the Maurer-Cartan equations:
\begin{equation}
d\theta^{L\,i} = \medio C_{jk}^i \theta^{L\,j}\wedge \theta^{L\,k}\,,
\end{equation}

\ni with analogous expression for the right-invariant counterpart, but
changing the sign to the structure constants, as before. These equations
state, for instance, that, for an Abelian group, all left- and right-invariant
1-forms are closed, and that left- and right-invariant 1-forms dual to
vector fields that are not in the commutant of $\calG$ are also closed.
These properties will be relevant below.

Let us consider a central extension \Gt of $G$ by $U(1)$ characterized by
a 2-cocycle $\xi:\ G\times G\rightarrow R$, which has to satisfy the
equations:
\begin{eqnarray}
\xi(g_1,g_2) + \xi(g_1*g_2,g_3) &=& \xi(g_1,g_2*g_3) + \xi(g_2,g_3) \nn\\
\xi(e,e)&=&0 \,,  \label{cociclo}
\end{eqnarray}

\ni for all $g_1,g_2,g_3\in G$, in order to define a (associative) group law.
This group law is given by:
\begin{eqnarray}
g'' &=& g'*g \nn \\
\z'' &=& \z'\z e^{i\xi(g',g)}\,,   \label{leyextendida}
\end{eqnarray}

\ni where $\z,\z',\z''\in U(1)$. Left- and right-invariant vector fields
for the extended group \Gt, denoted with a tilde, can be derived from the
ones of $G$ and from the 2-cocycle as follows:
\bea
\XL{i} &=& X^L_i +\derivparcial{\xi(g',g)}{g^i}|_{g=e,g'=g}\parcial{\phi} \nn\\
\XR{i} &=& X^R_i + \derivparcial{\xi(g',g)}{g'{}^i}|_{g'=e}\parcial{\phi}\,,
\eea

\ni where we have introduced $\z=e^{i\phi}$. Left- and right invariant 1-forms
do not change, and, of course, there are new left- and right invariant vectors
fields and 1-forms associated with the new variable $\z\in U(1)$. These are:
\bea
\XL{\z}&=&\parcial{\phi}=2{\rm Re}(i\z\parcial{\z})\equiv \Xi \nn \\
\XR{\z}&=& \Xi \nn \\
\theta^{L(\z)} &=& \frac{d\z}{i\z} +
            \derivparcial{\xi(g',g)}{g^i}|_{g'=g^{-1}} dg^i  \\
\theta^{R(\z)} &=& \frac{d\z}{i\z} +
            \derivparcial{\xi(g',g)}{g'^i}|_{g=g^{-1}} dg^i \,, \nn
\eea

 \ni where $\frac{d\z}{i\z}= d\phi$. We shall call
$\Theta\equiv \theta^{L(\z)}$ the Quantization 1-form. This
1-form defines a connection on the fibre bundle
$U(1)\rightarrow \Gtm \rightarrow G$, and will play an important role in our
formalism, since it contains all
the information about the dynamics of the system under study. In fact,
$\ThetaPC = \Theta-\frac{d\z}{i\z}$ is the Poincar\'e-Cartan 1-form, and
$d\Theta=d\ThetaPC$ is a presymplectic 2-form on $G$ which defines a
symplectic 2-form once the distribution generated by its kernel is removed.

Now let us assume that we add to $\xi$ the coboundary $\xi_{\lambda}$, 
generated by the function $\lambda$,
$\xi_{\lambda}(g',g)=\lambda(g'*g)-\lambda(g')- \lambda(g)$, with $\lambda$
satisfying $\lambda(e)=0$ for $\xi_{\lambda}$ to verify (\ref{cociclo}).
Then $\xi'=\xi + \xi_{\lambda}$ determines a new extended group
$\Gtm'$, and a new Quantization 1-form $\Theta'=\Theta+\Theta_{\lambda}$, with
\be
\Theta_{\lambda} = \lo_i \theta^{L\,i} - d\lambda \,.
\ee

The new presymplectic 2-form is $d\Theta'=d\Theta + d\Theta_{\lambda}$, with
$d\Theta_{\lambda}=\medio \lo_i C_{jk}^i \theta^{L\,j}\wedge \theta^{L\,k}$
(making use of the Maurer-Cartan equations).
We shall use this decomposition of $\Theta'$ and $d\Theta'$ to split an
arbitrary 2-cocycle $\xi'$ in the form
\be
\xi'=\xi + \xi_{\lambda}\,, \label{split}
\ee

\ni for some $\lambda(g)$. The term
$\xi$ is such that, when considered on its own, it determines a {\it pure
central extension}, i.e. a central extension for which the Lie algebra
satisfies: If $C_{ij}^{\z}\neq 0$, then $C_{ij}^k=0 \ \forall k\neq \z$.

 The term $\xi_{\lambda}$ is such that, when considered on its own, it 
determines a {\it pure pseudo-extension}, i.e a central extension for which 
the Lie algebra satisfies: $C_{ij}^{\z}=\lo_k C_{ij}^k, \ \forall i,j$, with 
$\vlo$ the gradient at the identity of $\lambda(g)$.

An arbitrary central extension determined by $\xi$ will belong to a given
cohomology class $[[\xi]]$ constituted by all 2-cocycles $\xi'$ differing
from $\xi$ by coboundaries with arbitrary generating functions $\lambda:\
G\rightarrow R$. This is the usual definition of the $2^{\rm nd}$
cohomology group $H^2(G,U(1))$ (see, for instance \cite{Bargmann}). Now we
are going to introduce subclasses $[\xi]$ inside $[[\xi]]$, called
pseudo-cohomology classes. For the sake of simplicity, we shall restrict to
the trivial cohomology class $[[\xi]]_0$ of 2-cocycles which admit a
generating function and are therefore coboundaries. The partition of
$[[\xi]]_0$ into pseudo-cohomology subclasses can be translated to any
other cohomology class using the decomposition (\ref{split}). The
equivalence relation defining the subclasses $[\xi]$ is given by:

\ni {\it Two coboundaries $\xi_\lambda$ and $\xi_{\lambda'}$ with
generating functions $\lambda$ and $\lambda'$, respectively, are in the
same subclass $[\xi]_{\vlo}$ if and only if
their gradients at the identity verify $\vlo{}'=Ad^*(g)\vlo$, for some
$g\in G$. }

In particular, if $\vlo{}'=\vlo$, $\xi_\lambda$ and $\xi_{\lambda'}$ are in the
same pseudo-cohomology class. This allows us always to choose representatives
that are linear in the canonical coordinates, $\xi_{\vlo}=\lo_i g^i$.

The condition $\vlo{}'=Ad^*(g)\vlo$ simply says that $\vlo{}'$ and $\vlo$
lie in the same coadjoint orbit in $\calG^*$, and it is justified because
$d\Theta_{\vlo{}'}$ and $d\Theta_{\vlo{}}$ are symplectomorphic, the
symplectomorphism being the pull-back of the coadjoint action
(see \cite{Marmo}).

The equivalence relation we have just introduced constitutes a partition of
the trivial cohomology class $[[\xi]]_0$ of coboundaries (or of any cohomology
class once translated by the relation (\ref{split})), but there is not a
one to one correspondence between pseudo-cohomology classes and coadjoint
orbits, since the coadjoint orbits must satisfy the integrality condition
(see \cite{Marmo}, and \cite{Pressley} for the proof) for $\xi_{\lambda}$
to define a central extension. This restriction can be expressed in a
different manner:

 The gradient at the identity $\vlo\in \calG^*$ defines a linear functional
of $\calG$ on $R$. But
it also defines a one-dimensional representation of the isotropy lie subalgebra
$\calG_{\vlo}$ of the point $\vlo$ under the coadjoint action of $G$ on $\calG^*$.
In particular, if $\vlo$ is invariant under
the coadjoint action
(i.e. it constitutes a zero dimensional coadjoint orbit), it defines a
one-dimensional representation of the whole Lie algebra $\calG$. The
condition of integrability
of the coadjoint orbit passing through $\vlo$ is nothing more than the
condition for $\vlo$ to be exponentiable (integrable) to a character of
the isotropy subgroup $G_{\vlo}$ (whose Lie algebra is $\calG_{\vlo}$).

The introduction of a pseudo-extension generated by $\lambda(g)$ in $G$,
defining a central extension \Gt, has
the effect of modifying left- and right-invariant vector fields in the following way:
\begin{equation}
\XL{i}=X^L_i + (X^L_i.\lambda - \lambda^0_i)\Xi\,\,,\qquad
\XR{i}=X^R_i + (X^R_i.\lambda - \lambda^0_i)\Xi\,.
\end{equation}

\ni It also modifies the commutation relations  in the Lie algebra
${\cal G}$ of $G$ (defining the commutation relations of $\tilde{\cal G}$):
%
\begin{equation}
[\XL{i},\XL{j}] = C_{ij}^k( \XL{k} + \lambda^0_k \Xi)\,,
\end{equation}

\ni  where $C_{ij}^k$ are the structure constants of the original algebra
${\cal G}$.
For right-invariant vector fields, we get the same commutation relations
up to a sign. Once the representations of \Gt have been obtained (using a technique like
GAQ, for instance), we recover the representations of $G$ by simply redefining the operators
(right-invariant vector fields) in the following manner:
\begin{equation}
\XR{i}\rightarrow \XR{i}{}'=\XR{i}+\lambda^0_i\Xi=X^R_i + (X^R_i.\lambda)\Xi\,.
\label{redefined-fields}
\end{equation}

It is trivial to check that the new generators $\XR{i}{}'$ satisfy the
(original) commutation relations of ${\cal G}$.

Once that the pseudo-extensions have been introduced and classified according to
equivalence classes, they can be treated as if they were {\it true extensions}
and the ordinary quantization techniques, in particular GAQ, can be applied.
We refer the reader to \cite{Marmo} for a detailed description of GAQ, and
here we shall simply use it to arrive at the irreducible representations of
$SL(2,\Real)$ in Sec. \ref{example}.

\section{Quasi-invariant measures}
\label{quasi-invariant}

For any Lie group  $G$, there exists a measure, the Haar
measure, which is invariant under the left or right
action of the group on itself. However,
if $M$ is a manifold on which there is a transitive action of $G$
(that is, $M$ is a homogenous space under $G$), the existence of an invariant
measure on $M$ is not guaranteed, despite that $M$ is locally
diffeomorphic to the quotient $G/H$ of $G$ by a certain closed subgroup $H$,
which is the isotropy group of an arbitrary point $x_0\in M$. More
precisely, each point in $M$ has a different isotropy group, although all of
them are conjugate to each other; in particular all are isomorphic.

It can be proven (see \cite{Mackey} and \cite{Barut}), however, that $M$
admits quasi-invariant measures. A measure
$d\mu(x)$ on $M$ is called quasi-invariant if $d\mu(gx)$ is equivalent to
$d\mu(x)$ for all $g\in G$, where $gx$ denotes the action of $G$ on $M$, and
the equivalence relation is defined among measures that have the same sets of
measure zero. Then the Radon-Nikodym theorem asserts that there exists a
positive function $\lambda$ (the Radon-Nikodym derivative) on $M$
such that $d\mu(gx)/d\mu(x)=\lambda(g,x)$.

Furthermore, it turns out that any two quasi-invariant measures are
equivalent (\cite{Barut,Mackey}). Therefore,
up to equivalence, there exists a unique quasi-invariant measure
$d\mu(x)$ with Radon-Nikodym derivative $\lambda(\cdot,x)$ on $M$.
The function $\lambda$ can be derived from a strictly positive, locally
integrable, Borel function $\rho(g)$
satisfying\footnote{Since we are considering the
quotient space $G/H$ instead of $H\backslash G$, i.e. we are changing left by right
with respect to \cite{Mackey,Barut}, modular functions get inverted.}
\begin{equation}
\rho(gh)= \frac{\Delta_G(h)}{\Delta_H(h)}\rho(g)\,, \label{rho-function}
\end{equation}

\ni where $\Delta_G, \Delta_H$ are the {\it modular} function of $G$ and
$H$, respectively (a modular function of $G$ is a non-negative functions on
$G$ such that, if $\mu_G(\cdot)$ is the left-invariant Haar measure on $G$,
then $\mu_G(R_g f) = \Delta(g)\mu_G(f)$, where $R_g$ means right
translation by the element $g$). A modular function is a homomorphism of
$G$ into the positive reals with the product as composition law). The
Radon-Nikodym derivative is given by:
\begin{equation}
\lambda(g,x)= \frac{\rho(gg')}{\rho(g')}\,, \label{lambda-function}
\end{equation}

\ni where $g'$ is any element whose image under the natural projection
$G\rightarrow M$ is $x$. This definition makes sense since
$\frac{\rho(gg')}{\rho(g')}$ depends only on $x$ and not on the particular
choice of $g'$.

Note that if $\Delta_H(h)=\Delta_G(h),\,\forall h\in H$, then $\rho(gh)=\rho(g)$, so that
we can choose $\rho(g)=1$ and $\lambda(g,x)=1$ as the Radon-Nikodym
derivative. Thus, in this case, $M$ admits an invariant measure under $G$.

Let us rewrite the above considerations in infinitesimal terms. Defining
the {\it modular constants}
$k^G_i\equiv \frac{\partial\Delta_G(g)}{\partial g^i}|_{g=e},\,i=1,\ldots,n={\rm dim}G$,
and similarly for $k^H_i,\, i=1,\ldots,p={\rm dim}H$, we can rephrase the
$\rho$-function condition (\ref{rho-function}) as:
\begin{equation}
X^L_{i}\rho(g) = k^{G/H}_i\rho(g)\,, \label{inf-rho-function}
\end{equation}

\ni where $k^{G/H}_i\equiv k^G_i-k^H_i,\,i=1,\ldots,p$. Modular constants possess
properties derived from those of modular functions. Firstly, it can be proven that
$k^G_i=\sum_{j=1}^{n} C_{ij}^j$, and accordingly,
$k^{G/H}_i=\sum_{j=p+1}^{n} C_{ij}^j$, where we have assumed that the
first $p={\rm dim}H$ elements of $\calG$ belong to $\calH$, the Lie algebra
of $H$. In addition,
$k^G_i,\,i=1,\ldots,n$ define a character $k^G$ of the Lie algebra ${\cal G}$ of $G$,
coming from the fact that $\Delta_G(g)$ defines a character of $G$, in such a way that
$k^G(X^L_{i})=k^G_i$. This property implies linearity, and also
$C_{ij}^l k^G_l=0$, since $k^G([X^L_{i},X^L_{j}])=0$. As a result, $k^G_i=0$ for $G$
semisimple.

However, $k^H_i$ can be non-trivial, even if $H$ is a subgroup of a semisimple group $G$,
allowing for non-trivial $k^{G/H}_i$, and, according to (\ref{inf-rho-function}), for the
possibility of homogeneous spaces with non-invariant, although quasi-invariant, measures.

Let us develop a constructive technique for obtaining quasi-invariant
measures on homogeneous spaces. That is, a procedure for constructing
$\rho$-functions satisfying (\ref{rho-function}) (or
(\ref{inf-rho-function})). According to Mackey \cite{Mackey}, such a function always exists,
although the proof of his theorem is not constructive.

Consider the left-invariant Haar measure $\Omega^L$ on $G$. This is an n-form,
with $n={\rm dim}G$, and can be written, up to a constant, as
$\Omega^L=\theta^{L\,1}\wedge\theta^{L\,2}\wedge \cdots \wedge\theta^{L\,n}$, 
where $\theta^{L\,i},\,i=1,\ldots,n$, is the set of left invariant 1-forms on 
$G$ dual to a given basis $\{X^L_{i}\}$ of left-invariant vector fields. Let 
us suppose that the first $p={\rm dim}H$ elements in these bases correspond to
left-invariant 1-forms and vector fields of $H$, respectively. Then we 
tentatively define a measure on
$G/H$ as:
\begin{equation}
\Omega^L_H = i_{X^L_{p}}i_{X^L_{p-1}} \ldots i_{X^L_{1}}\Omega^L =
\theta^{L\,p+1}\wedge\ldots\wedge\theta^{L\,n}\,.
\end{equation}

In general, $\Omega^L_H$ is not an invariant measure on $G/H$; in fact, it is 
not even a measure on $G/H$, in the sense that it does not fall down to the 
quotient. This can
be checked by computing its invariance properties under $X^L_{i},\,i=1,\ldots,p$. After a
few computations we get $L_{X^L_{i}}\Omega^L_H = - k^{G/H}_i \Omega^L_H$. Therefore,
if $k^{G/H}_i\neq 0$ for some $i$, $\Omega^L_H$ does not fall down to the quotient, and
this is the same condition for $G/H$ not to have a strictly invariant measure. Therefore,
these two facts seem to be related. Indeed, if we look for a function $\rho$ on $G$ such
that $L_{X^L_{i}}(\rho\Omega^L_H )=0,\,i=1,\ldots,p$, we find that $\rho$ has to be a
$\rho$-function, satisfying $X^L_{i}\rho= k^{G/H}_i\rho$, as in (\ref{inf-rho-function}).

Now we have to prove that equation (\ref{inf-rho-function}) always has non-trivial 
solutions.
We know from Mackey \cite{Mackey}, that equation (\ref{rho-function}) always has a solution,
but we would like to provide a proof in infinitesimal terms and, moreover, we would like
to construct  the solutions explicitly.

Let us consider the Radical of $\calH$,  $\Rad\calH$ -- the maximal solvable ideal of $\calH$.
We know that $\calH/\Rad\calH$ is semisimple. According to the previous considerations,
the $k_i$'s  vanish on this quotient. Thus, the non-trivial $k_i$'s lie only on $\Rad\calH$, which
is solvable. According to one of Lie's theorems \cite{Humphreys}, a solvable algebra of
operators always possesses a common eigenvector. We proceed to construct it as follows:

Let us consider the equation $X^L_{i}\rho = k^{G/H}_i\rho,\,i=1,\ldots,p$. Let $\chi$ be the
general solution of $X^L_{i}\chi=0$, which always exists and which we know how to construct,
according to the Frobenious theorem.
Then we can write $\rho=\chi h$, where $h$ is a particular solution  of
$X^L_{i}h = k^{G/H}_i h$, with $X^L_{i}\in \Rad\calH$ (the rest of the equations give zero,
and since $h$ is a particular solution, we can choose so as not to depend on the corresponding
variables). Then Lie's Theorem guarantees the existence of such a function $h$,
since $\Rad\calH$ is solvable.

Once we have constructed the measure $\rho\:\Omega^L_H$ on $G/H$, we must check its
invariance properties under the action of $G$. For this, we compute
$L_{X^R_{i}}(\rho\:\Omega^L_H)=\frac{1}{\rho}(X^R_{i}.\rho)(\rho\:\Omega^L_H),\,i=1,\ldots,n$.
The result is that $\rho\:\Omega^L_H$ is quasi-invariant under $G$ and the divergence of the
vertor field $X^R_{i}$ is $\frac{1}{\rho}(X^R_{i}.\rho)$. Once the divergence of all vector
fields have been computed, it is very easy to modify the (infinitesimal) action of the
group $G$ in order to restore the invariance of $\rho\:\Omega^L_H$, by defining the new
vector fields:
\begin{equation}
\XR{i} = X^R_i + \frac{1}{2\rho}(X^R_i.\rho)\,,
\end{equation}

\ni i.e., right-invariant vector fields are modified with the addition of a
multiplicative term, half the divergence of the corresponding vector field. In
the context of Sec. \ref{pseudoextensiones}, we could think of this
redefinition as coming from a pseudo-extension of $G$ by means of some
pseudo-cocycle generated by a certain function
$\lambda$ on $G$. In fact, this is the case, since the extra term can be
written as $X^R_i(\medio \log\rho)$, i.e., the function $\lambda$, according
to equation (\ref{redefined-fields}), would be $\lambda=-i\medio\log\rho$.
Note the presence of the imaginary constant $i$ in $\lambda$ (so that
$\lambda$ is a pure imaginary function) revealing that $G$ has been centrally
pseudo-extended by $R^+$ instead of $U(1)$. Therefore, the invariance of a
measure on a quotient space $G/H$ can be restored by means of a central
extension of $G$ by $R^+$ with generating function $-i\medio\log\rho$, where
$\rho$ is a $\rho$-function.

If we compute the commutation relations of the redefined vector fields, we
get:
\begin{equation}
[\XR{i},\XR{j}] = - C_{ij}^k\XR{k}\, ,
\end{equation}

\ni showing that this pseudo-extension does not modify the commutation
relations. As in Sec. \ref{pseudoextensiones}, we can compute the gradient of
the generating function $\lambda$ at the identity, proving to be
$\lambda^0_i=-\frac{i}{2} k^{G/H}_i,\,i=1,\ldots,n$. It is pure imaginary,
as would be expected of a pseudo-extension by $R^+$.

\section{Example: Representations of $SL(2,\Real)$}
\label{example}

Let us consider, as an example of application of the formalism 
developed above, the study of the unitary and irreducible
representations of $G=SL(2,\Real)$. Since this group is non-simply connected,
in order to apply our previous considerations, we shall consider its
universal covering group $\bar{G}$, with $p:\bar{G}\rightarrow SL(2,\Real)$ the
covering map, which is a group homomorphism. The kernel of $p$ is $Z$, the
first homotopy group of $SL(2,\Real)$. It is easy to check that a unirrep $U$
of $\bar{G}$ is also a unirrep of $SL(2,\Real)$ if and only if Ker$p$ is
represented as phases, i.e $U(g)=e^{i\alpha_g},\,\forall g\in\,{\rm ker}p$.
Therefore, we shall compute the representations $U$ of $\bar{G}$ and then
retain only the ones that verify $U(g)=e^{i\alpha_g},\,\alpha_g\in R,\,\forall
g\in\,{\rm Ker}p$. For simplicity, we shall denote $\bar{G}$ just by $G$,
bearing in mind that at the end we wish to get the representations of
$SL(2,\Real)$.

Since $SL(2,\Real)$ is semisimple, it has no non-trivial central extensions by
$U(1)$; i.e. its second cohomology group $H^2(G,U(1))=\{e\}$. However, as
shown in Sec. \ref{pseudoextensiones}, this group admits non-trivial
pseudo-extensions by $U(1)$, which can be classified into pseudo-cohomology
classes. These pseudo-cohomology classes are in one-to-one correspondence
with the coadjoint orbits of $SL(2,\Real)$ with integral symplectic 2-form (see
\cite{Marmo}).

 Thus, we must first study the coadjoint orbits of $SL(2,\Real)$. These can be
classified
into three types: the 1-sheet hyperboloids, the 2-sheets hyperboloids, and
the cones. The cones are really three different orbits, the upper and lower
cones and the origin. The origin is the only zero-dimensional orbit, and is
associated with the only one-dimensional representation (character) of
$SL(2,\Real)$, the trivial one.

 As we shall see below, the 1-sheet hyperboloids are associated with the
Principal continuous series of unirreps of $SL(2,\Real)$, the 2-sheet hyperboloids
are associated with the Principal discrete series of unirreps and the two
cones are associated with the Mock representations.

\subsection{The group law}

 The $SL(2,\Real)$ group can be parameterized by:
 \begin{equation}
 SL(2,\Real) = \left\{ \left( \begin{array}{cc} a&b \\ c&d \end{array} \right)\in
M_2(\Real)\;/\; ad-bc=1 \right\}\,.
 \end{equation}

 If $a\neq 0$ (the case $a=0$ is treated in an analogous manner, changing $a$ by $c$),
 we can eliminate $d$, $d=\frac{1+bc}{a}$, and we arrive at the  following group law from
 matrix multiplication:
 \begin{eqnarray}
 a'' &=& a'a + b'c \nn\\
 b'' &=& a'b + b'\frac{1+bc}{a} \\
 c'' &=& c'a + \frac{1+b'c'}{a'}c\,. \nn
 \end{eqnarray}

Left- and right-invariant vector fields are easily derived from the group law:
\begin{equation}
\begin{array}{lcl}
X^L_a &=& a\parcial{a} + c\parcial{c} - b\parcial{b} \\
X^L_b &=& a\parcial{b} \\
X^L_c &=& \frac{1+bc}{a} \parcial{c} + b\parcial{a}
\end{array} \qquad
\begin{array}{lcl}
X^R_a &=& a\parcial{a} + b\parcial{b} - c\parcial{c} \\
X^R_b &=& \frac{1+bc}{a} \parcial{b} + c\parcial{a}\\
X^R_c &=& a\parcial{c}\,.
\end{array}
\end{equation}

The Lie algebra satisfied by the (say, left-invariant) vector fields is:
\begin{eqnarray}
[X^L_a,X^L_b] &=& 2 X^L_b \nn\\
{}[X^L_a,X^L_c] &=& -2 X^L_c \\
{}[X^L_b,X^L_c] &=& X^L_a \,, \nn
\end{eqnarray}

\ni and the Casimir for this Lie algebra is given by
$\hat{C}=\medio(X^L_a)^2 + X^L_bX^L_c+ X^L_cX^L_b$. The left-invariant 1-forms (dual to the
set of left-invariant vector fields) are given by:
\begin{eqnarray}
\theta^{L(a)} &=& \frac{1+bc}{a} da -bdc \nn\\
\theta^{L(b)} &=& \frac{1}{a}db -\frac{b^2}{a}dc + \frac{b}{a} \frac{1+bc}{a} da  \\
\theta^{L(c)} &=& adc -cda\,. \nn
\end{eqnarray}

The exterior product of all left-invariant 1-forms constitutes a (left-invariant)
volume form on the whole group (Haar measure):
\begin{equation}
\Omega^L = \theta^{L(a)}\wedge \theta^{L(b)} \wedge \theta^{L(c)}=
\frac{1}{a} da\wedge db\wedge dc\,.
\end{equation}

\subsection{Pseudo-extensions}
\label{pseudo-extensiones-SL(2,R)}

The different (classes of) pseudo-extensions of $SL(2,\Real)$ by $U(1)$ are
classified, according to the discussion in Sec. \ref{pseudoextensiones}, by
the coadjoints orbits of the group $SL(2,\Real)$. Let us parameterize ${\cal G}^*$
by $\{\alpha,\beta,\gamma\}$, a coordinate system associated with the base
$\{X^L_a,X^L_b,X^L_c\}$ of ${\cal G}$. Instead of looking for the different
coadjoint orbits by direct computation, we can classify them by means of the
Casimir functions. The Casimirs $C_i$ are invariant functions under the
coadjoint action of the group on ${\cal G}^*$, so that the equations $C_i=c_i$
define hypersurfaces on ${\cal G}^*$ invariant under the coadjoint action. Of
course, these hypersurfaces could be the union of two or more coadjoint
orbits, and we shall need extra conditions to characterize them (these are
called invariant relations, see \cite{Marmo,Levi-Civita}).

 The only (independent) Casimir function for $SL(2,\Real)$ is
$C=\medio\alpha^2+\beta\gamma$. This is a quadratic function, and therefore its
level sets are conic sections.

It is more appropriate for our purposes to perform the change of variables
$\alpha=\alpha\,,\,\beta= \mu+\nu\,,\,\gamma=\mu-\nu$. In terms of the new
variables, the Casimir function is written $C=\medio\alpha^2+2\mu^2-2\nu^2$.
In this form, it is easy to identify the conics, of which there are essentially three
types, depending on whether $C>0,\,C=0$ or $C<0$. The case $C>0$ corresponds
to 1-sheet hyperboloids; the case $C=0$ corresponds to the two cones and the
origin, i.e. the union of three coadjoint orbits; and finally, the case $C<0$
corresponds to 2-sheets hyperboloids (i.e. the union of two coadjoint orbits).

 Now we select a particular point $\vlo$ in each coadjoint orbit,
which will be used to define a pseudo-extension in $SL(2,\Real)$ (different
choices of $\vlo$ in the same coadjoint orbit will lead to equivalent
pseudo-extensions). For the case
$C>0$, the easiest choice is $\vlo=(\alpha,0,0)$. For $C=0$,
we have $\vlo=(0,0,0)$ for the origin, and we can choose $\vlo=(0,0,\gamma<0)$
for the upper cone and $\vlo=(0,0,\gamma>0)$ for the lower cone.
Finally, for the case $C<0$, we select $\vlo=(0,\beta>0,\gamma=-\beta)$
for the upper sheet and $\vlo=(0,\beta<0,\gamma=-\beta)$ for the lower
sheet of the 2-sheets hyperboloid.

\subsection{Representations associated with the 1-sheet hyperboloid: Principal
Continuous Series}

 According to the above discussion, let us choose $\vlo=(\alpha,0,0)$ as
the representative point in the 1-sheet hyperboloids. We need to look for a
function $\lambda$ on $SL(2,\Real)$ satisfying
$\parcial{g^i}{\lambda(g)}|_{g=e}=\lambda^0_i$. The easiest one would be a
function linear on the coordinate $a$, but we should take into account
that $a$ is not a canonical coordinate, since its composition law is
multiplicative. That is, the uniparametric subgroup associated with it is
$R^+$ instead of $R$ (the value of $a$ at the identity of the group is $1$
instead of $0$). Thus, we can select for $\lambda(g)=\alpha \log a$ or rather
$\lambda(g)=\alpha (a-1)$, since the generating function $\lambda$ must
satisfy $\lambda(e)=0$ for $\xi_{\lambda}$ to satisfy (\ref{cociclo}).

 Let us fix $\lambda(g)=\alpha(a-1)$, to be precise (the other choice would
lead to en equivalent result). The representation achieved when applying
GAQ to the resulting group will be associated with the coadjoint orbit for
which the Casimir is $C=\medio \alpha^2>0$. The resulting group law for
$SL(2,\Real)$ pseudo-extended by $U(1)$ by means of the two-cocycle
$\xi_\lambda$ is:
\begin{eqnarray}
a'' &=& a'a + b'c \nn\\
b'' &=& a'b + b'\frac{1+bc}{a} \nn\\
c '' &=& c'a + \frac{1+b'c'}{a'}c \\
\z'' &=& \z'\z e^{i\alpha(a'a+b'c-a'-a+1)} \,.\nn
\end{eqnarray}

 Left- and right-invariant vector field, obtained as usual from the group law,
are:
\begin{equation}
\begin{array}{lcl}
\XL{a} &=& a\parcial{a} + c\parcial{c} - b\parcial{b} +\alpha(a-1)\Xi\\
\XL{b} &=& a\parcial{b} \\
\XL{c} &=& \frac{1+bc}{a} \parcial{c} + b\parcial{a} + \alpha b\Xi \\
\XL{\z} &=& \parcial{\phi}=2{\rm Re}(i\z\parcial{\z})\equiv \Xi
\end{array} \qquad
\begin{array}{lcl}
\XR{a} &=& a\parcial{a} + b\parcial{b} - c\parcial{c} +\alpha(a-1)\Xi\\
\XR{b} &=& \frac{1+bc}{a} \parcial{b} + c\parcial{a}+ \alpha c\Xi\\
\XR{c} &=& a\parcial{c} \\
\XR{\z} &=& \Xi\,.
\end{array}
\end{equation}

Left- and right-invariant 1-forms associated with the variables of $SL(2,\Real)$
remain the same, and there are extra left- and right-invariant 1-forms
associated with the variable $\z$. We are interested in the left-invariant
one, which is:
\begin{equation}
\Theta \equiv \theta^{L(\z)} = \frac{d\z}{i\z} +
                \alpha( \theta^{L(a)}-da) = \frac{d\z}{i\z} +
                \alpha(\frac{1+bc-a}{a}da - bdc)\,.
\end{equation}

The resulting Lie algebra is that of $SL(2,\Real)$ with one of the commutators
modified:
\begin{eqnarray}
[\XL{a},\XL{b}] &=& 2 \XL{b} \nn\\
{}[\XL{a},\XL{c}] &=& -2 \XL{c} \\
{}[\XL{b},\XL{c}] &=& \XL{a} + \alpha\Xi\,. \nn
\end{eqnarray}

The 2-form
\begin{equation}
d\Theta = \alpha ( dc\wedge db + \frac{c}{a} db\wedge da +
                   \frac{b}{a}dc\wedge da )
\end{equation}

\ni defines a presymplectic structure on \Gt. The characteristic module, or
more precisely, ker$\:d\Theta\cap$ker$\Theta$, is generated by the
{\bf characteristic subalgebra}, ${\cal G}_C=<\XL{a}>$. We should remember that the
characteristic subalgebra is nothing more than the isotropy subalgebra
${\cal G}_{\vlo}$ of the point $\vlo\in\calG$.

 Now we have to look for polarization subalgebras.  These should contain
the characteristic subalgebra ${\cal G}_C$ and must be horizontal (i.e., in the
kernel of $\Theta$). There are essentially two, and these lead to unitarily
equivalent representations (since they are related by the adjoint action
of the Lie algebra on itself, and this turns out to be a unitary
transformation). We shall choose as polarization
\begin{equation}
{\cal P}=< \XL{a},\XL{b}>\,,
\end{equation}

\ni and this, by solving the equation $\XL{a}\Psi=\XL{b}\Psi=0$, provides the
wave functions
$\Psi=\z e^{-i\alpha(\kappa-1)}\kappa^{i\alpha}\Phi(\tau)$, where
$\kappa\equiv a$ and $\tau\equiv \frac{c}{a}$. The action of the
right-invariant vector fields on polarized wave functions is:
\begin{eqnarray}
\XR{a} \Psi &=& \z e^{-i\alpha(\kappa-1)}\kappa^{i\alpha}
[-2\tau\frac{d\ }{d\tau}]\Phi(\tau) \nn\\
\XR{b} \Psi &=& \z e^{-i\alpha(\kappa-1)}\kappa^{i\alpha}
[i\alpha\tau -\tau^2\frac{d\ }{d\tau}]\Phi(\tau) \\
\XR{c} \Psi &=& \z e^{-i\alpha(\kappa-1)}\kappa^{i\alpha}
[\frac{d\ }{d\tau}]\Phi(\tau)\,. \nn
\end{eqnarray}

According to Sec. \ref{pseudoextensiones}, the right-invariant generators
should be redefined as
$\XR{g^i} \rightarrow \XR{g^i}{}'=\XR{g^i}+\lambda^0_i\Xi$ in order to obtain
the representations of $G$,
and this affects only the generators $\XR{a}$, which changes to
$\XR{a}{}'=\XR{a} + \alpha\Xi$. Its action on polarized wave functions turns
out to be:
\begin{equation}
\XR{a}{}' \Psi = \z e^{-i\alpha(\kappa-1)}\kappa^{i\alpha}
[i\alpha-2\tau\frac{d\ }{d\tau}]\Phi(\tau)\,.
\end{equation}

The representation of $SL(2,\Real)$ here constructed is irreducible but not
unitary. One way of viewing it (before discussing 
integration measures) is to consider the Casimir operator, which is the
quadratic operator
$\hat{C}=\medio (\XR{a})^2 + \XR{b}\XR{a} + \XR{c}\XR{b}$. After the
pseudoextension and redefinition of operators ($\XR{a}$ should be
changed by $\XR{a}{}'$), the resulting Casimir operator, $\hat{C}{}'$, acts on
polarized wave functions as $\hat{C}{}'\Psi=(-\alpha^2/2 + i\alpha)\Psi$. The
fact that it is a number
reveals that the representation is irreducible, but since it is not real, the
representation cannot be unitary (the Casimir is a quadratic function of
(anti-)Hermitian operators, and  should therefore be a self-adjoint operator
in any unitary representation).

 The reason for this lack of unitarity is that the support manifold
for the representation does not admit an invariant measure.
Since the process of polarizing wave functions
really amounts to reducing the space of functions to those
defined in the quotient $G/G_{\cal P}$, where $G_{\cal P}$ is the group
associated with the polarization subalgebra ${\cal P}$), the support manifold
is given by $G/G_{\cal P}$, which is naturally
an homogeneous space under $G$. According to Sec. \ref{quasi-invariant},
it may well happen that $G/G_{\cal P}$ does not admit an invariant measure,
and in fact this is the case. However,
the existence of quasi-invariant measures is granted, and this fact will allow
us to restore the unitarity of the representation.

If we compute the measure on $G/G_{\cal P}$, derived from the left Haar measure
$\Omega^L$ on $G$,
we obtain $\Omega^L_{\cal P} = i_{\XL{b}} i_{\XL{a}}\Omega^L= adc - cda$. When
expressed  in terms of the new variables $\kappa$ and $\tau$, it takes the form
$\Omega^L_{\cal P} = \kappa^2 d\tau$. Taking into account that $G/G_{\cal P}$
is  parameterized by $\tau$, now becomes clear why the representation is not
unitary: the measure does not even fall down  to the quotient.


A solution to this problem consists in choosing any quasi-invariant measure on
$G/G_{\cal P}$ and  introducing the appropriate Radon-Nikodym derivative
\cite{Mackey,Barut}. Here, we propose another, yet equivalent, solution to this
lack of unitarity, giving a new insight into the problem according to Sec.
\ref{quasi-invariant}. We shall consider a pseudo-extension of $G$ by $R^+$,
rather than $U(1)$. The reason is that we wish to restore the unitarity
of a non-unitary representation, and for this we need a ``piece" of
non-unitary representation, in such a way that the resulting representation is
unitary. To enable a direct comparison with the treatment of
Mackey, we shall employ the equivalent technique of non-horizontal
polarizations instead of that of pseudo-extensions. A non-horizontal
polarization $\calP^{\rm n.h.}$ is a polarization in which the horizontality
condition has been relaxed. The polarization equations acquire  the form:
$\XL{j}\Psi=i\alpha_j\Psi,\,\forall \XL{j}\in\calP^{\rm n.h.}$ (see
\cite{Marmo} for a discussion on the equivalence between pseudo-extensions and
non-horizontal polarizations).

The key point is to keep $\Omega^L_{\cal P}$ as the measure on $G/G_{\cal P}$,
and to impose the polarization conditions
$\XL{i}\tilde{\Psi}=\medio k^{G/G_{\cal P}}_i\tilde{\Psi}$, instead of
$\XL{i}\Psi=0,\,\forall \XL{i}\in{\cal P}$. In finite terms, this condition is
written as:
\begin{equation}
\tilde{\Psi}(g*h) = \sqrt{ \frac{\Delta_G(h)}{\Delta_H(h)}}\, \tilde{\Psi}(g)\,.
\end{equation}

We can rephrase this by saying that $\tilde{\Psi}$ is a
$\medio$-$\rho$-function\footnote{Note that, according to Sec.
\ref{quasi-invariant}, the generating function for the
pseudo-extension by $R^+$ would be $\lambda=-\frac{i}{2}\log\rho=-i\log\rho^{\medio}$,
with $\lambda^0_i=-\frac{i}{2}k^{G/G_\calP}_i=\alpha_i$.}. The purpose of
this definition is to make $\tilde{\Psi}^*\tilde{\Psi}'$ a $\rho$-function,
with  two $\medio$-$\rho$-functions $\tilde{\Psi}$ and $\tilde{\Psi'}$, in
such a way that $\tilde{\Psi}^*\tilde{\Psi}'\Omega^L_{\cal P}$ is a
well-defined quantity on $G/G_{\cal P}$ and can be integrated with respect to
$\tau$. In other words, $\tilde{\Psi}^*\tilde{\Psi}'$ is a
$\rho$-function necessary to make $\Omega^L_{\cal P}$ a quasi-invariant
measure on $G/G_{\cal P}$.

To begin, we must compute the modular constants
$k^{G/G_{\cal P}}_i=k^G_i-k^{G_{\cal P}}_i,\,i=1,\ldots,p$. Firstly, since $G=SL(2,\Real)$ is semi-simple,
$k^G_i=0,\,i=1,\ldots,n$. Secondly, we have $k^{G_{\cal P}}_a=2$ and $k^{G_{\cal P}}_b=0$.
Therefore, $k^{G/G_{\cal P}}_a=-2$ and $k^{G/G_{\cal P}}_b=0$.

Accordingly, the new polarization equations we have to solve are:
\begin{equation}
\XL{a}\tilde{\Psi} = -\tilde{\Psi}\,\,,\qquad \XL{b}\tilde{\Psi}=0\,.
\end{equation}

It is easy to verify that the solutions of these new polarization equations are of the form:
\begin{equation}
\tilde{\Psi}(g)=a^{-1}\Psi(g)\,,
\end{equation}

\ni where $\Psi(g)$ is a solution of the previous (horizontal) polarization equations. Thus,
the form of the solutions is:
\begin{equation}
\tilde{\Psi} = \z \kappa^{-1} e^{-i\alpha(\kappa-1)}\kappa^{i\alpha}\Phi(\tau)\,.
\end{equation}

Now it it clear why
$\tilde{\Psi}^*\tilde{\Psi}'\Omega^L_{\cal P}=\Phi(\tau)^*\Phi'(\tau)d\tau$
can be integrated in $G/G_{\cal P}$; the $\kappa$ dependence has been removed.

The right-invariant vector fields, when acting on $\medio$-$\rho$-functions,
acquire extra terms that restore the unitarity of the representation\footnote{The
difference between pseudo-extensions and non-horizontal polarizations lie in
the fact that pseudo-extensions modify the left- and right-invariant vector
fields and non-horizontal polarizations modify
the wave functions. The extra term in the reduced operators is a consequence
of their acting on modified wave functions.}:
\begin{equation}
\XR{i}\tilde{\Psi} = \kappa^{-1}\XR{i}\Psi + \kappa^{-1}(\kappa\XR{i}.\kappa^{-1})\Psi \,.
\end{equation}

In this way, the final representation has the form, restricted to its action on $\Phi(\tau)$:
\begin{eqnarray}
\XR{a}{}' \Phi(\tau) &=& [-1 + i\alpha-2\tau\frac{d\ }{d\tau}]\Phi(\tau) \nn\\
\XR{b} \Phi(\tau) &=&[-\tau + i\alpha\tau -\tau^2\frac{d\ }{d\tau}]\Phi(\tau) \\
\XR{c} \Phi(\tau) &=& [\frac{d\ }{d\tau}]\Phi(\tau) \,.\nn
\end{eqnarray}

W can readly verify that these operators are self-adjoint with respect to the
quasi-invariant measure $d\tau$ (what remains of $\Omega^L_{\cal P}$ after
multiplication by the factor $\kappa^{-2}$
contained in the wave functions). Even more, the Casimir operator, acting on the new
wave functions, turns out to be real, revealing that the representation is now unitary:
\begin{equation}
\hat{C}'\Phi(\tau) = -\medio (1+\alpha^2)\Phi(\tau)\,.
\end{equation}


\subsection{Representations associated with the cones: Mock representation}

In accordance with Sec. \ref{pseudo-extensiones-SL(2,R)}, let us choose $\vlo=(0,0,\gamma)$ as
the representative point in the cone. If $\gamma<0$ we are in the upper cone and if
$\gamma>0$ we are in the lower cone. We have to look for a
function $\lambda$ on $SL(2,\Real)$ satisfying
$\parcial{g^i}{\lambda(g)}|_{g=e}=\lambda^0_i$. The easiest one is the function
linear on the coordinate $c$, since here $c$ is a true canonical coordinate, 
and therefore, we fix $\lambda(g)=\gamma c$.

 The representation obtained when applying
GAQ to the resulting group will be associated with one of the coadjoint orbit for
which the Casimir is $C=0$. The resulting group law for
$SL(2,\Real)$ pseudo-extended by $U(1)$ by means of the two-cocycle
$\xi_\lambda$ is:
\begin{eqnarray}
a'' &=& a'a + b'c \nn\\
b'' &=& a'b + b'\frac{1+bc}{a} \nn\\
c '' &=& c'a + \frac{1+b'c'}{a'}c \\
\z'' &=& \z'\z e^{i\gamma(c'a+\frac{1+b'c'}{a'}c-c'-c)} \,.\nn
\end{eqnarray}

 Left- and right-invariant vector field, derived as usual from the group law,
are:
\begin{equation}
\begin{array}{lcl}
\XL{a} &=& a\parcial{a} + c\parcial{c} - b\parcial{b} +\gamma c\Xi\\
\XL{b} &=& a\parcial{b} \\
\XL{c} &=& \frac{1+bc}{a} \parcial{c} + b\parcial{a} + \gamma (\frac{1+bc}{a}-1) \Xi \\
\XL{\z} &=& \parcial{\phi}=2{\rm Re}(i\z\parcial{\z})\equiv \Xi
\end{array} \qquad
\begin{array}{lcl}
\XR{a} &=& a\parcial{a} + b\parcial{b} - c\parcial{c} -\gamma c\Xi\\
\XR{b} &=& \frac{1+bc}{a} \parcial{b} + c\parcial{a}\\
\XR{c} &=& a\parcial{c} + \gamma(a-1)\Xi\\
\XR{\z} &=& \Xi\,.
\end{array}
\end{equation}

The left-invariant 1-form associated with the variable $\z$ is:
\begin{equation}
\Theta \equiv \theta^{L(\z)} = \frac{d\z}{i\z} +
                \gamma( \theta^{L(c)}-dc) = \frac{d\z}{i\z} +
                \gamma((a-1)dc -cda)\,.
\end{equation}

The resulting Lie algebra is, again, that of $SL(2,\Real)$ with one of the commutators
modified, in this case the one giving $\XL{c}$ on the r.h.s.:
\begin{eqnarray}
[\XL{a},\XL{b}] &=& 2 \XL{b} \nn\\
{}[\XL{a},\XL{c}] &=& -2 (\XL{c}+\gamma \Xi) \\
{}[\XL{b},\XL{c}] &=& \XL{a}\,. \nn
\end{eqnarray}

The 2-form
\begin{equation}
d\Theta = 2\gamma da\wedge dc
\end{equation}

\ni defines a presymplectic structure on \Gt. The characteristic subalgebra
is ${\cal G}_C=<\XL{b}>$.
In this case, there is essentially one polarization, given by:
\begin{equation}
{\cal P}=< \XL{b},\XL{a}>\,,
\end{equation}

\ni and this provides, by solving the equation $\XL{a}\Psi=\XL{b}\Psi=0$, the
wave functions $\Psi=\z e^{-i\gamma c}\Phi(\tau)$, where again
$\tau\equiv \frac{c}{a}$. The action of
right-invariant vector fields on polarized wave functions is:
\begin{eqnarray}
\XR{a} \Psi &=& \z e^{-i\gamma c}[-2\tau\frac{d\ }{d\tau}]\Phi(\tau) \nn\\
\XR{b} \Psi &=& \z e^{-i\gamma c}[-\tau^2\frac{d\ }{d\tau}]\Phi(\tau) \\
\XR{c} \Psi &=& \z e^{-i\gamma c}[\frac{d\ }{d\tau} -i\gamma]\Phi(\tau)\,. \nn
\end{eqnarray}

 The redefinition of the right-invariant generators
$\XR{g^i} \rightarrow \XR{g^i}{}'=\XR{g^i}+\lambda^0_i\Xi$ in order to obtain the
representation of $G$,
affects only to the $\XR{c}$ generator, which changes to
$\XR{c}{}'=\XR{a} + \gamma\Xi$. Its action on polarized wave functions turns
out to be:
\begin{equation}
\XR{c}{}' \Psi = \z e^{-i\gamma c}[\frac{d\ }{d\tau}]\Phi(\tau)\,.
\end{equation}

The representation of $SL(2,\Real)$ here constructed, as in the case of the 1-sheet
hyperboloid, is irreducible but not unitary.

 The reason for this lack of unitarity is the same as before, that is, the lack of an
invariant measure on the support manifold $G/G_{\cal P}$.
In fact, the polarization ${\cal P}$ is the same as in the case of the
1-sheet hyperboloid, only the vector fields are slightly different, since
they come from different pseudo-extensions. Therefore, the wave functions are
essentially the same as before, and consequently $G/G_{\cal P}$ is the same
as in the case of the 1-sheet hyperboloid.

The measure on $G/G_{\cal P}$ is again
$\Omega^L_{\cal P} = i_{\XL{b}} i_{\XL{a}}\Omega^L= adc - cda = \kappa^2
d\tau$, which does not fall down  to the quotient.

Thus, we keep $\Omega^L_{\cal P}$ as the measure on $G/G_{\cal P}$, and we
impose the polarization conditions
$\XL{i}\tilde{\Psi}=\medio k^{G/G_{\cal P}}_i\tilde{\Psi}$, instead of
$\XL{i}\Psi=0,\,\forall \XL{i}\in{\cal P}$. In other words, we impose
$\tilde{\Psi}$ to be a $\medio$-$\rho$-function in such a way that
$\tilde{\Psi}^*\tilde{\Psi}'$ is a $\rho$-function,  $\tilde{\Psi}$ and
$\tilde{\Psi'}$ being two $\medio$-$\rho$-functions. Now,
$\tilde{\Psi}^*\tilde{\Psi}'\Omega^L_{\cal P}$ is a well-defined quantity on
$G/G_{\cal P}$ and can be integrated with respect to $\tau$.

Modular constants
$k^{G/G_{\cal P}}_i=k^G_i-k^{G_{\cal P}}_i,\,i=1,\ldots,p$, are the same as
before, since $G_{\cal P}$ is the same group. Therefore,
$k^{G/G_{\cal P}}_a=-2$ and $k^{G/G_{\cal P}}_b=0$.

The new polarization equations are:
\begin{equation}
\XL{a}\tilde{\Psi} = -\tilde{\Psi}\,\,,\qquad \XL{b}\tilde{\Psi}=0\,.
\end{equation}

\noindent with solutions:
\begin{equation}
\tilde{\Psi}(g)=a^{-1}\Psi(g)\,,
\end{equation}

\ni where $\Psi(g)$ is a solution of the previous (horizontal) polarization equations. Thus,
the form of the solutions is:
\begin{equation}
\tilde{\Psi} = \z \kappa^{-1} e^{-i\gamma \kappa \tau}\Phi(\tau)\,.
\end{equation}


The right-invariant vector fields, when acting on $\medio$-$\rho$-functions, acquire extra
terms restoring the unitarity of the representation:
\begin{equation}
\XR{i}\tilde{\Psi} = \kappa^{-1}\XR{i}\Psi + \kappa^{-1}(\kappa\XR{i}.\kappa^{-1})\Psi \,.
\end{equation}

\ni This way, the final representation restricted to its action on
$\Phi(\tau)$ has the form :
\begin{eqnarray}
\XR{a}\Phi(\tau) &=& [-1 -2\tau\frac{d\ }{d\tau}]\Phi(\tau) \nn\\
\XR{b} \Phi(\tau) &=&[-\tau -\tau^2\frac{d\ }{d\tau}]\Phi(\tau) \\
\XR{c}{}' \Phi(\tau) &=& [\frac{d\ }{d\tau}]\Phi(\tau) \,.\nn
\end{eqnarray}

Again, we can readily verify that these operators are self-adjoint with respect
to the quasi-invariant measure $d\tau$ (what remains of $\Omega^L_{\cal P}$
after multiplication by the factor $\kappa^{-2}$ contained in the wave
functions). Therefore, the representation is now unitary.

This representation can be seen as the limit $\alpha\rightarrow 0$ of the Principal series
of representations. We should stress at this point that the representation does not depend
on $\gamma$, nor even on its sign. Therefore, we obtain the same representation for both
cones, which are clearly equivalent. The reason for this equivalence is that the
group isomorphism $(a,b,c)\rightarrow (a,b,-c)$ induces a unitary transformation
between the two representations. This representation (up to equivalence) is 
called the
Mock representation and is associated with the two cones.


\subsection{Representations associated with the 2-sheets hyperboloids: Discrete Series}

 According to Sec. \ref{pseudo-extensiones-SL(2,R)}, we can choose
the point $\vlo=(0,\beta>0,\gamma=-\beta)$ in the upper sheet and
$\vlo=(0,\beta<0,\gamma=-\beta)$ in the lower sheet of the 2-sheets hyperboloid, to define
the pseudo-extension of $SL(2,\Real)$ by $U(1)$. Let us consider
$\vlo=(0,\beta,-\beta)$, keeping the sign of $\beta$ undetermined for the time being.

The easiest
function $\lambda$ on $SL(2,\Real)$ satisfying
$\parcial{g^i}{\lambda(g)}|_{g=e}=\lambda^0_i$ is the function
linear on the coordinate $(b-c)$, since here $b$ and $c$ are true canonical coordinates.
Therefore, we fix $\lambda(g)=\beta (b-c)$.

  The representation obtained when applying
GAQ to the resulting group will be associated with one of the coadjoint orbits for
which the Casimir is $C=-\beta^2<0$. The resulting group law for
$SL(2,\Real)$, pseudo-extended by $U(1)$ by means of the two-cocycle
$\xi_\lambda$, is:
\begin{eqnarray}
a'' &=& a'a + b'c \nn\\
b'' &=& a'b + b'\frac{1+bc}{a} \nn\\
c '' &=& c'a + \frac{1+b'c'}{a'}c \\
\z'' &=& \z'\z e^{i\beta((a'-1)b-(a-1)c'+ \frac{1+bc-a}{a}b'-
                  \frac{1+b'c'-a'}{a'}c)} \,.\nn
\end{eqnarray}

 Left- and right-invariant vector field are:
\begin{equation}
\begin{array}{lcl}
\XL{a} &=& a\parcial{a} + c\parcial{c} - b\parcial{b} -\beta(b+c)\Xi\\
\XL{b} &=& a\parcial{b} +\beta(a-1)\Xi \\
\XL{c} &=& \frac{1+bc}{a} \parcial{c} + b\parcial{a} - \beta (\frac{1+bc-a}{a}) \Xi \\
\XL{\z} &=& \Xi
\end{array} \qquad
\begin{array}{lcl}
\XR{a} &=& a\parcial{a} + b\parcial{b} - c\parcial{c} +\beta(b+c)\Xi\\
\XR{b} &=& \frac{1+bc}{a} \parcial{b} + c\parcial{a} +\beta(\frac{1+bc-a}{a})\Xi\\
\XR{c} &=& a\parcial{c} - \beta(a-1)\Xi\\
\XR{\z} &=& \Xi\,.
\end{array}
\end{equation}

The left-invariant 1-form associated with the variable $\z$ is:
\begin{eqnarray}
\Theta \equiv \theta^{L(\z)} &=& \frac{d\z}{i\z} +
                \beta( \theta^{L(b)}-db - \theta^{L(c)}+dc) =
                \frac{d\z}{i\z} +
                \beta\left[\frac{1-a}{a}db - (1+a+\frac{b^2}{a})dc + \right.\nn\\
            & &  \left.  (\frac{b}{a^2}(1+bc)-c)da \right]\,.
\end{eqnarray}

The resulting Lie algebra is, as in the other cases, the one of $SL(2,\Real)$
with some of the commutators modified, in this case those giving $\XL{b}$
and $\XL{c}$ on the r.h.s.:
\begin{eqnarray}
[\XL{a},\XL{b}] &=& 2 (\XL{b}+\beta\Xi) \nn\\
{}[\XL{a},\XL{c}] &=& -2 (\XL{c}-\beta \Xi) \\
{}[\XL{b},\XL{c}] &=& \XL{a}\,. \nn
\end{eqnarray}

The 2-form defining a presymplectic structure on \Gt is
\begin{equation}
d\Theta = -2\beta\left[\frac{b}{a}db\wedge dc +\frac{1+bc}{a^2} da\wedge db
+ da\wedge dc\right]\,.
\end{equation}

The characteristic subalgebra turns out to be ${\cal G}_C=<\XL{b}-\XL{c}>$.
Looking for a polarization subalgebra containing the characteristic
subalgebra, we get into trouble, since there is no such real subalgebra.
We are forced to complexify the algebra, and then we find (essentially)
two complex polarizations:
\begin{equation}
{\cal P}=< \XL{b}-\XL{c}, \XL{b}+\XL{c}\pm i\XL{a} >\,.
\label{pola-compleja}
\end{equation}

Clearly, the solution to these polarization equations are complex
functions defined on a complex submanifold of the complexification of
$SL(2,\Real)$. These will be holomorphic or anti-holomorphic, depending on the
choice of sign in (\ref{pola-compleja}). The explicit construction of the
representations in the discrete series, according to the group quantization
framework, was firstly given in Ref. \cite{A23} in connection to the quantum
dynamics of a free particle on Anti-de Sitter space-time. Higher-order, real
polarizations were used in Ref. \cite{Oscilata} in the study of the
relativistic harmonic oscillator. They have also been considered in conformal
field theory as factor of $SO(2,2)\approx SL(2,\Real)\otimes SL(2,\Real)$
representations \cite{Vacuum}.

\section*{Acknowledgements}
We thank G. Marmo for very useful comments on Sec.
\ref{quasi-invariant}.


\begin{thebibliography}{99}

\bibitem{JMP23}   Aldaya, V. and de Azc\'{a}rraga, J.: J. Math. Phys. {\bf 23},
                    1297 (1982)

\bibitem{Comm1}   Aldaya, V. Navarro-Salas, J. and  Ram\'{\i}rez, A.:
                  Commun. Math. Phys. {\bf 121}, 541-556 (1989)

\bibitem{frachall} Aldaya, V., Calixto, M. and Guerrero, J.: Commun. Math.
                   Phys. {\bf 178}, 399-424 (1996)

\bibitem{Kirillov} Kirillov, A.A.: {\it Elements of the Theory of
                  Representations}, Springer-Verlag (1976)

\bibitem{Kostant} Kostant, B.: {\it Quantization and Unitary Representations},
                  Lecture Notes in Math. {\bf 170}, Springer-Verlag, Berlin
                  (1970)

\bibitem{virazorro} Aldaya, V. and Navarro-Salas, J.: Commun. Math. Phys.
                    {\bf 139}, 433 (1991)

\bibitem{Witten}  Witten, E.: Commun. Math. Phys.: {\bf 114}, 1 (1988)

\bibitem{pseudo}  Aldaya, V. and  de Azc\'{a}rraga, J.A.: Int. J. Theo. Phys.
                 {\bf 24}, 141 (1985)

\bibitem{Marmo}  Aldaya, V., Guerrero, J. and Marmo, G.: Int. J. Mod. Phys.
               {\bf A12}, 3 (1997)

\bibitem{Woodhouse} Woodhouse, N.: {\it Geometric Quantization}, Oxford
                    University Press (1980)

\bibitem{Bargmann}  Bargmann, V.: Ann. Math. {\bf 59}, 1 (1954)

\bibitem{Pressley}  Pressley A., and  Segal, G.: {\it Loop Groups},
                  Clarendon Press, Oxford (1986)

\bibitem{Mackey} G.W. Mackey: Ann. Math. {\bf 55}, 101 (1952)

\bibitem{Barut}  Barut, A.O. and  Raczka, R. {\it Theory of Group
                 Representations and Applications}, World Scientific
                 Publishing, Singapore (1986)


\bibitem{Humphreys} Humphreys, J.E.: {\it Introduction to Lie Algebras and
                    Representation Theory}, Graduate Texts in Math.
                    Springer-Verlag (1972)


\bibitem{Levi-Civita} Levi-Civita, T. and Amaldi, U.: {\it Lezioni di
       Meccanica Razionale}, Zanichelli, Bologna (1974) (reprinted version of
       1949 edition)

\bibitem{A23} Aldaya, V., Azc\'arraga, J.A. de, Bisquert, J. and Cerver\'o,
              J.M.: J. Phys. {\bf A23}, 707 (1990)

\bibitem{Oscilata} Aldaya, V., Bisquert, J., Guerrero, J. and Navarro-Salas,
             J.: Rep. Math. Phys. {\bf 37}, 387 (1996)

\bibitem{Vacuum} Aldaya, V., Calixto, M. and Cerver\'o, J.M.: Commun. Math.
               Phys. {\bf 200}, 325 (1999)

\end{thebibliography}
\end{document}